\documentclass[conference,letterpaper]{IEEEtran}
\IEEEoverridecommandlockouts 

\usepackage{cite}
\usepackage[pdftex]{graphicx}   
\graphicspath{{./figs/}}
\usepackage[cmex10]{amsmath}
\usepackage{array}
\usepackage{amsfonts, amssymb}
\usepackage{amscd}
\usepackage{fixltx2e} 
\usepackage{url}
\usepackage{enumerate}

\newtheorem{defi}{Definition}

\newcommand{\Z}{{\mathbb{Z}}}
\newcommand{\zz}{\Z_2\Z_4}

\newcommand{\F}{{\mathbb{F}}}

\newcommand{\dd}{\displaystyle}

\newcommand{\codi}{{\cal C}}
\newcommand{\HH}{{\cal H}}

\newcommand{\bx}{\mathbf{x}}
\newcommand{\bs}{\mathbf{s}}

\newcommand{\bh}{\mathbf{h}}

\newcommand{\bw}{\mathbf{w}}

\newcommand{\by}{\mathbf{y}}

\newcommand{\additive}{\Z_2\Z_4}

\newcommand{\dos}{{\mathbf{2}}}

\begin{document}

\title{Product Perfect $\additive$-linear codes in Steganography\thanks{This work was partially supported by the Spanish MICINN Grants MTM2009-08435, PCI2006-A7-0616, and also by the \textit{Comissionat per a Universitats i Recerca del DIUE de la Generalitat de Catalunya} and the \textit{European Social Fund} with Grants 2009SGR1224 and FI-DGR.}}

\author{\IEEEauthorblockN{Josep Rif\`{a} and
Lorena Ronquillo}
\IEEEauthorblockA{Department of Information and Communications Engineering,\\
Universitat Aut\`{o}noma de Barcelona,\\
08193-Cerdanyola del Vall\`{e}s, Spain.\\
Email: Josep.Rifa@autonoma.edu, Lorena.Ronquillo@autonoma.edu}
}

\maketitle

\begin{abstract}
Product perfect codes have been proven to enhance the performance of the $F5$ steganographic method, whereas perfect $\zz$-linear codes have been recently introduced as an efficient way to embed data, conforming to the $\pm 1$-steganography. In this paper, we present two steganographic methods. On the one hand, a generalization of product perfect codes is made. On the other hand, this generalization is applied to perfect $\zz$-linear codes. Finally, the performance of the proposed methods is evaluated and compared with those of the aforementioned schemes.
\end{abstract}


\section{Introduction}\label{sec:intro}

Digital steganography is an information hiding application which consists of hiding data within a commonly used media in such a way that unintended recipients are not only unable to detect the presence of embedded data, but also given no reason for suspecting that anything is unusual. This is the main difference from encryption, which only prevents the adversary from decoding the message and not from suspecting that a secret message is being sent.

{\it Matrix encoding} is a steganographic method introduced by Crandall~\cite{cra} and analyzed by Bierbrauer et al.~\cite{bier}, which requires the sender and the recipient to agree in advance on a parity check matrix $H$, and the secret message is then extracted by the recipient as the syndrome (with respect to $H$) of the received cover object. This method was made popular by Westfeld~\cite{west}, who incorporated a specific implementation using Hamming codes. The resulting method is known as the $F5$ algorithm, and it can embed $t$ bits of message in $2^t - 1$ cover symbols by changing, at most, one of them.

\medskip
The following two parameters are used to evaluate the performance of a
steganographic method over a cover message of $N$ symbols: the \textit{average distortion} $D=\frac{R_a}{N}$, where $R_a$
is the expected number of changes over uniformly distributed messages; and the
\textit{embedding rate} $E= \frac{t}{N}$, which is the amount of bits that can be
hidden in a cover message. Given two methods with the same embedding rate, the one with smaller average distortion is better. Following the terminology used by Fridrich et al.~\cite{fri}, the tuple $(D,E)$ will be called {\it $CI$-rate}.

As Willems et al. in~\cite{WiMa}, we will also assume that a discrete source produces a sequence $\bx=(x_1,\ldots,x_N)$, where $N$ is the block length, $x_i \in \aleph =\{0,1,\ldots,2^B-1\}$, and $B\in \{8,12,16\}$ depends on the kind of source (digital image, CD audio, etc). Let $\bs \in \{1,\ldots,M\}$ be the message we want to hide into a host sequence $\bx$, which produces a composite sequence $\by = f(\bx,\bs)$, for $\by =(y_1,\ldots y_N)$ and $y_i\in \aleph$. The sequence $\by$ is obtained from distorting $\bx$, and that distortion will be assumed to be of squared-error type (see~\cite{WiMa}). In these conditions, information can be carried by the least significant bit (LSB) or by the two least significant bits of each $x_i$. An appropriate solution for the first case comes from applying the $F5$ algorithm~\cite{west}, which has been improved in~\cite{riri} by using the Kronecker product of the corresponding generator matrices of two binary perfect codes. The latter case is known as ``$\pm 1$-steganography" and the magnitude of changes is limited to $1$, that is, $y_i=x_i +c$, where $c\in \{0,+1,-1\}$. This case has usually involved the use of ternary codes~\cite{WiMa,fri} until the results from~\cite{ririro}, which introduces a method based on perfect $\zz$-linear codes. This kind of codes are not linear but have a representation using a parity check matrix that makes them as efficient as the Hamming codes. The steganographic method therein presented not only outperforms the one obtained by direct sum of ternary Hamming codes, but it also deals better with the {\it extreme grayscale values problem}. That is, the problem we may have when the steganographic method requires adding one unit to a grayscale value which already has the maximum allowed value $2^B-1$, or substracting one unit from a grayscale of value $0$.

Let $H_q(x)=\frac{1}{\log_2(q)}(H(x)+x\log_2(q-1))$ be the $q$-ary entropy function~\cite{bier1} on the interval $[0,(q-1)/q]$, where $H(x)=-x\log_2(x)-(1-x)\log_2(1-x)$ is the usual binary entropy function on the interval $[0,1/2]$. We call {\it normalized embedding rate} the ratio $e=\frac{{H_q}^{-1}(E)}{D}$, where ${H_q}^{-1}(\cdot)$ is the inverse of the $q$-ary entropy function $H_q(x)$. In the binary case, $e$ will be computed by considering the binary entropy function $H_2(x)$, whereas in the $\pm 1$-steganography the ternary entropy function $H_3(x)$ is used. One of the purposes of steganographic methods is to approach the upper bound on the normalized embedding rate $e$ subject to the constraint of an average distortion $D$. This upper bound on $e$ for a fixed $D$ is $e \leq 1$, and it is the same for any kind of steganography, be it binary or $\pm 1$-steganography.

In this paper we propose a technique based on products of perfect $\zz$-linear codes and compare its performance with that of the product binary perfect codes~\cite{riri} and perfect $\zz$-linear codes~\cite{ririro}.

The current paper has been organized as follows. Some basic concepts on perfect $\zz$-linear codes, as well as the steganographic method based on these codes~\cite{ririro}, are reviewed in Section~\ref{sec:z2z4stegano}. Then, Section~\ref{sec:ppcGeneral} reviews the product perfect codes method~\cite{riri} and presents a generalization that enhances its performance. In Section~\ref{sec:productz2z4}, this generalization is used for $\zz$-additive perfect codes in $\pm 1$-steganography. Finally, the paper is concluded in Section~\ref{sec:conclusions}.

\section{Perfect $\additive$-linear codes and steganography}\label{sec:z2z4stegano}

Any non-empty subgroup $\codi$ of $\Z_2^\alpha \times \Z_4^\beta$ is a {\it $\zz$-additive code}, where $\Z_2^\alpha$ denotes the set of all binary vectors of length $\alpha$ and $\Z_4^\beta$ is the set of all quaternary vectors of length $\beta$. Let $\phi$ be the usual {\it Gray map} from $\Z_4$ onto $\Z_2^2$, where $\phi(0)=(0,0)$, $\phi(1)=(0,1)$, $\phi(2)=(1,1)$, and $\phi(3)=(1,0)$; and let $\Phi: \Z_2^{\alpha}\times\Z_4^{\beta} \longrightarrow \Z_2^{n}$ be the {\it extended Gray map} given by $$\Phi(u_1, \ldots, u_{\alpha} | v_1, \ldots, v_{\beta})=(u_1, \ldots, u_{\alpha} | \phi(v_1), \ldots, \phi(v_{\beta})).$$

A $\zz$-additive code $\codi$ is isomorphic to an abelian structure like
$\Z_2^{\gamma}\times \Z_4^{\delta}$. Therefore, $\codi$ has $|\codi|=2^\gamma 4^\delta $ codewords, where $2^{\gamma+\delta}$ of them are of order two. We call such code $\codi$ a
{\it $\zz$-additive code of type $(\alpha,\beta;\gamma,\delta)$} and its
binary image $C=\Phi(\codi)$ is a {\it $\zz$-linear code of type
$(\alpha,\beta;\gamma,\delta)$}. Note that the Lee distance of a $\zz$-additive code $\codi$ coincides with the Hamming distance of the $\zz$-linear code $C$, and that the binary code $C$ may not be linear.

The {\it $\Z_2\Z_4$-additive dual code} of $\codi$, denoted by ${\cal
C}^\perp$, is defined as the set of vectors in $\Z_2^\alpha \times \Z_4^\beta$ that are orthogonal to every codeword in $\codi$, being the definition of inner product in $\Z_2^{\alpha}\times \Z_4^{\beta}$ the following (see~\cite{z2z4}):
\begin{equation} \label{inner}
  \langle u,v \rangle=2(\sum_{i=1}^{\alpha}
  u_iv_i)+\sum_{j=\alpha+1}^{\alpha+\beta}
u_jv_j\in \Z_4, \end{equation}
 where $u,v\in \Z_2^{\alpha}\times \Z_4^{\beta}$ and computations are made considering the zeros and ones in the $\alpha$ binary coordinates as quaternary zeros and ones, respectively.

The binary code $C_\perp=\Phi({\cal C}^\perp)$, of length $n=\alpha+2\beta$, is called the {\it
$\Z_2\Z_4$-dual code} of $C$.

A $\zz$-additive code $\codi$ is said to be {\it perfect} if code $C=\Phi(\codi)$ is a perfect $\zz$-linear code, that is all vectors in $\Z_2^{n}$ are within distance one from a codeword and the distance between two codewords is, at least, $3$.

It is well known~\cite{br} that for any $m\geq 2$ and each $\delta$ $\in$ $\{0,\ldots,\lfloor \frac{m}{2}\rfloor \}$ there exists a perfect $\zz$-linear code $C$ of binary length $n=2^m-1$, such that its $\zz$-dual code is of type $(\alpha,\beta;\gamma,\delta)$, where $\alpha=2^{m-\delta}-1$, $\beta=2^{m-1}-2^{m-\delta-1}$ and $\gamma=m-2\delta$ (note that the binary length can be computed as $n=\alpha+2\beta$). This allows us to write the parity check matrix $\HH_{\codi}$ of any $\zz$-additive perfect code $\codi$ for a given value of $\delta$. Matrix $\HH_{\codi}$ can be represented by taking as columns all possible vectors in $\Z_2^\gamma \times \Z_4^\delta$, up to sign changes. In this representation, there are $\alpha$ columns which correspond to the binary part of codewords in $\codi$, and $\beta$ columns of order four which correspond to the quaternary part. We agree on a representation of the $\alpha$ binary coordinates as coordinates in $\{0,2\} \in \Z_4$. Let $\bh_i$, for $i \in \{1,\ldots,\alpha+\beta \}$, denote the $i$-th column vector of $\HH_{\codi}$.

\bigskip

Now we proceed to review how a perfect $\zz$-linear code $C=\Phi(\codi)$ can be used in steganography. Consider its $\zz$-dual, of type $(\alpha,\beta;\gamma,\delta)$, which gives us a parity check matrix $\HH_{\codi}$ with $\gamma$ rows of order two and $\delta$ rows of order four.

Take $N=2^{m-1}$ and let $x=(x_1,\ldots,x_N)$ be a source of grayscale symbols such that $x_i \in \aleph=\{0,1,\ldots,2^B-1\}$, where, for instance, $B=8$ for grayscale images.

We assume each grayscale symbol $x_i$ is represented as a binary vector $(v_{(B-1)i},\ldots,v_{1i},v_{0i})$, obtained by first representing $x_i$ in base $4$ and then applying the Gray map $\phi$ to every quaternary symbol in the base $4$ representation. For example, the grayscale value $239$ is represented as the quaternary vector $(3233)$, which then gives rise to the binary vector $(10111010)$ after applying the Gray map $\phi$.

The $N$-length packet $x$ of symbols is translated into a vector $\bw$ of $\alpha$ binary and $\beta$ quaternary coordinates. The binary coordinates come from taking the least significant bit of the binary representation of $x_1$, that is $v_{01}$, along with the two least significant bits $v_{1i},v_{0i}$ of the following $(\alpha+1)/2$ grayscale symbols $x_i$. The quaternary coordinates of $\bw$ come from taking the two least significant bits of the last $\beta$ symbols $x_i$ and interpreting them as integer numbers $\phi^{-1}(v_{1i},v_{0i})$ in $\Z_4$.

The obtained vector $\bw \in \Z_2^{\alpha} \times \Z_4^{\beta}$ is then distorted according to the matrix encoding method~\cite{cra,west} in such a way that $\HH_{\codi}{\bw}^T + \epsilon \cdot \bh_i=\bs$ holds, where $\bs$ $\in$ $\Z_2^{\gamma}\times \Z_4^{\delta}$ is the secret message we want to embed in $x$, the value of $\epsilon$ can be $\{0,1,3\}$, the syndrome vector of $\bw$ with respect to the parity check matrix $\HH_{\codi}$ is $\HH_{\codi}{\bw}^T$, and $\bh_i$ is a column vector in $\HH_{\codi}$.
This method also deals with the extreme grayscale values in a rather efficient way: when a symbol $x_i$ having an extreme value, be it $0$ or $2^B-1$, has to be distorted in a way that would lead its value out of the range defined by $\aleph$, two other symbols are changed instead one magnitude. One of these symbols is always $x_1$.
This method has $CI$-rate $\dd (D_m,E_m) = \left ( \frac{2N-1 + \frac{N-1}{2^{B-2}}}{2N^2} , \frac{1+\log(N)}{N} \right ) $. We refer the reader to~\cite{ririro} for further details on this steganographic scheme.

\section{Product of perfect codes and steganography}\label{sec:ppcGeneral}

Let $\F_q$ be a finite field of $q$ elements, where $q$ is a prime power. Let $C$ be a Hamming code over $\F_q$ of length $n=\dd\frac{q^m-1}{q-1}$ and dimension $n-m$. Let $G_C$, $H_C$ be, respectively, a generator matrix and a parity check matrix for $C$.

\begin{defi}
The Kronecker product of two matrices $A=[a_{r,t}]$ and $B =[b_{i,j}]$ over
$\F_q$ is a new matrix $A \otimes B$ obtained by changing any element
$a_{r,t}$ in $A$ by the matrix $a_{r,t} B$.
\end{defi}

The $q$-ary code $C^{\perp}_{H_C \otimes H_C}$, that is the dual of the code constructed by taking $H_C \otimes H_C$ as generator matrix, is a $[n^2,n^2-m^2]$ code  with covering radius $\rho_{C^{\perp}_{H_C \otimes H_C}}=m$ (see~\cite{crc}). Codewords of $C^{\perp}_{H_C \otimes H_C}$ can be seen as $n\times n$ matrices whose rows or columns are codewords in the $q$-ary Hamming code $C$.

The $q$-ary code $C_{G_C \otimes G_C}$, constructed from the generator matrix $G_C \otimes G_C$ is a $[n^2,(n-m)^2]$ code with covering radius $\rho_{C_{G_C \otimes G_C}}=n+1+2(n-m-1)$~\cite{crc}, whose codewords can be seen as $n\times n$ matrices where both rows and columns are codewords in the $q$-ary Hamming code $C$. We will refer to code $C_{G_C \otimes G_C}$ as {\it product code}.

There is an efficient steganographic method~\cite{riri} which uses the above defined code $C_{G_C \otimes G_C}$, for $q=2$, to embed data. Given two binary Hamming codes of the same length $n=2^m-1$, their product is considered, which gives a linear code of length $n=(2^m-1)^2$ and dimension $(n-m)^2$. Its codewords can be seen as $(2^m-1)\times(2^m-1)$ matrices where every row and every column are codewords in the binary Hamming code.
The embedding scheme therefore consists of first taking blocks in the cover source of size $(2^m-1)\times(2^m-1)$, and then applying the $F5$ algorithm to every row and also to the first $c$ columns, for $1 \leq c \leq 2^{m-1}-1$. As proved in~\cite{riri}, the performance one can obtain with this method is better than the one obtained by just using the conventional $F5$ algorithm on the corresponding codes with the same average distortion. We refer to~\cite{riri} for further details on this method.

\bigskip

Now, in this paper we will proceed with a generalization of the above procedure, by taking the product of more than two $q$-ary Hamming codes.

Let $C$ be a $q$-ary Hamming code of length $n=\frac{q^m-1}{q-1}$, dimension $n-m$, with generator matrix $G_C$ and parity check matrix $H_C$, as defined at the beginning of this section. Take the code $C^{'}$ of all the $n\times n$ matrices such that all their rows, as well as their first column, are codewords in the $q$-ary Hamming code $C$. Code $C^{'}$ is a $[n^2,n(n-m)-m]$ code, $C_{G_C \otimes G_C} \subset C^{'} \subset C^{\perp}_{H_C \otimes H_C}$ with covering radius $\rho_{C^{'}}=n+1$.

For the sake of a well understanding, the following reasoning will be limited to the binary case. However, a generalization to the $q$-ary case is straightforward.

Just as the method based on the product of two Hamming codes from~\cite{riri}, this procedure consists of a row embedding and a column embedding steps. We will
take the LSB bit of every grayscale symbol in the cover source and form blocks of size $n\times n$, where $n=2^m-1$.
Let $c_{i,j}$ be the coordinate in the $i$-th row and $j$-th column of these blocks, where $i,j \in \{1,\ldots,n\}$.

\medskip

\begin{enumerate}[1)]

\item \underline{Rows Embedding:}

\medskip

The matrix encoding standard procedure~\cite{cra,west} applied to every row lets us embed $\frac{m}{n}$ bits with an average distortion of $\frac{1}{n+1}$ coordinates, thus giving a $CI$-rate of $(\frac{1}{n+1},\frac{m}{n})$.

\medskip

\item \underline{Column Embedding:}

\medskip

After processing all rows, we can embed $\frac{m}{n}$ additional bits with an average distortion of $\frac{1}{n+1}$ by applying the same standard procedure to the first column.

However, note that the following situations can happen when processing this column:

\begin{itemize}
\item No coordinate needs to be changed in $\frac{1}{n+1}$ cases because the first column may already have, by chance, the desired value.
\item We may need to change a coordinate $c_{i,1}$ in $\frac{n}{n+1}$ cases. In this case, the $i$-th row may have been already modified in the corresponding row embedding with a probability of $\frac{n}{n+1}$, while it may have not been modified with a probability of $\frac{1}{n+1}$.

Let us consider the $i$-th row was modified in the coordinate $c_{i,j}$, for $j>1$. In this case, we will also have to restore the original value of $c_{i,j}$ and distort another appropiate coordinate $c_{i,k}$, for $k \in \{2,\ldots,n\}$ and $k \neq j$, such that the distortion being introduced now by the column embedding is compensated and does not affect the embedding in the $i$-th row (see Lemma 2 from~\cite{riri}). Note that this situation is also including the case in which the coordinate that was modified during the $i$-th row embedding is precisely the coordinate $c_{i,1}$ we now need to change to embed data in the column. In summary, if the $i$-th row was modified, no matter in which coordinate, the column embedding step will introduce one distortion besides the ones introduced by the row embedding step. 

\medskip

Otherwise, if during the column embedding we need to distort a coordinate $c_{i,1}$ and the $i$-th row was not modified, then we will also need to distort two more coordinates within the same row, $c_{i,j}$ and $c_{i,k}$, for $j,k \in \{2,\ldots,n\}$ and $j \neq k$, to make up for this distortion. Hence, the column embedding step will be now introducing three changes.

\end{itemize}

\end{enumerate}

In short, we can leave invariant the average distortion of the row embedding step, but $\frac{(n+3)/(n+1)}{n+1}$ should be added for the embedding in the first column. Note that this is only a tight upper bound on the average distortion, as we will later show.

\bigskip

By the method just described we can embed $m$ bits into the first column and also in every row of the matrix; therefore, we embed $(n+1)m$ bits in $n^2$ coordinates. The average distortion is upper bounded by $\frac{(n+3)/(n+1)}{n+1}$ for the coordinates in the first column and $\frac{1}{n+1}$ in each of the $n$ rows. Summing this up, the  average distortion is bounded by $\frac{n\frac{(n+3)/(n+1)}{n+1}+n^2\frac{1}{n+1}}{n^2}=\frac{1}{n+1} \left (1+\frac{(n+3)/(n+1)}{n} \right )$.

The method we propose in the present paper consists of repeating over and over the same procedure. Hence, we can generalize the computations of the average distortion and the embedding rate by using ${G_C}^l=G_C \otimes (G_C \otimes \cdots \otimes G_C)$. In each step ${G_C}^l=G_C \otimes {G_C}^{l-1}$, only the first column in the first component $G_C$ will be used to embed information.

Let $D_l$ be the average distortion at the $l$-th step. As computed before, we have $D_1=\frac{1}{n+1}$ and $D_2=\frac{1}{n+1}(1+\frac{(n+3)/(n+1)}{n})$. In the general case we have:
$$
D_l=\frac{1}{n+1} + \xi D_{l-1},
$$
where $\xi = \frac{n+3}{n(n+1)}$.

Now, the overall average distortion can be computed as $ \frac{1}{n+1}\big(1+\xi+\cdots +\xi^{l-1}\big)$, which converges asymptotically very fast to 
$$\frac{1}{n+1} \left ( \frac{\xi^{l}-1}{\xi-1} \right ) \rightarrow \frac{1}{n+1} \left ( \frac{1}{1-\xi} \right ) = \frac{1}{n+1} \left ( \frac{n(n+1)}{n^2-3} \right ).$$
As for the embedding rate, it can be computed as $\frac{(1+n+n^2+\cdots +n^{l-1})m}{n^l}$, which converges to
$$\frac{(1+n+n^2+\cdots +n^{l-1})m}{n^l} = m\frac{\frac{n^l-1}{n-1}}{n^l}\rightarrow \frac{m}{n-1}.$$

Finally, we obtain the asymptotical $CI$-rate $\left ( \frac{n}{n^2-3},\frac{m}{n-1} \right )$.

\medskip

Note that we are not able to generate an embedding scheme for any $CI$-rate but only for natural values of $m$. However, given any non-allowable parameter $D$ for the average distortion, we can always take two codes with $CI$-rates $(D_1,E_1)$ and $(D_2,E_2)$, where $D_1 < D < D_2$, such that their direct sum gives rise to a new $CI$-rate $(D,E)$, with $D=\lambda D_1+(1-\lambda) D_2$ and $E=\lambda E_1+(1-\lambda) E_2$.

\begin{figure}[htb!]
\centering 
\includegraphics[scale=0.45]{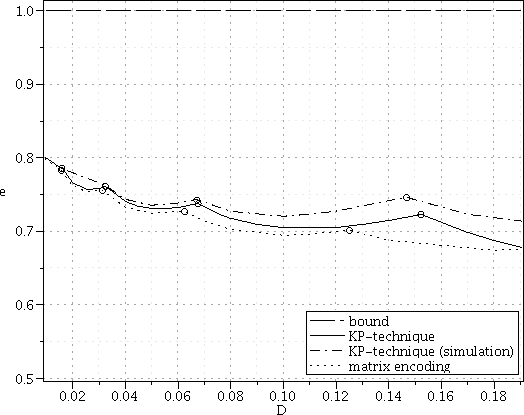}
\caption{Normalized embedding rate $e$ as a function of the average distortion $D$, of steganographic methods based on the matrix encoding procedure~\cite{cra,west}, and on the Kronecker product technique, using an upper bound on the average distortion ("KP-technique") and using the experimental results ("KP-technique (simulation)").}\label{fig:graphicKronecker}
\end{figure}

A comparison of the normalized embedding rate $e=\frac{{H_q}^{-1}(E)}{D}$, where $q=2$, as a function of the average distortion $D$ for the introduced Kronecker product technique (KP-technique) and the standard matrix encoding procedures~\cite{cra,west} is shown in \figurename~\ref{fig:graphicKronecker}. As explained before, this plot has been made by first computing the allowable points $(D,e)$, and then applying the direct sum between the codes corresponding to two contiguous points $(D_1, e_1)$ and $(D_2,e_2)$, where $D_1 < D_2$.

\medskip

For the sake of simplicity, some particular cases which may produce a lower distortion have been omitted in the computation of the distortion in the above $CI$-rate. For this reason, the distortion $D$ in that $CI$-rate is an upper bound.
As an example of one of these cases, recall that the column embedding step of our procedure is introducing three distortions when we need to change the $c_{i,1}$ coordinate and the $i$-th row was not modified in the row embedding step. Note, however, that there is one particular case in which we may need to introduce two distortions instead of three. This happens when there exist two other coordinates in the same column, $c_{j,1}$ and $c_{k,1}$, for $j,k \in \{1,\ldots,n\}$ and $j \neq k$, whose distortion is equivalent to distort only $c_{i,1}$, and both $j$-th and $k$-th rows were modified in the row embedding step. It is easy to see that we can distort coordinates $c_{j,1}$ and $c_{k,1}$ instead of $c_{i,1}$, and perform afterwards the appropiate changes to compensate these distortions in the $j$-th and $k$-th rows, respectively. Therefore, the column embedding step will be introducing two distortions besides the ones introduced in the row embedding step, and not three, as we previously stated. However, if no two other coordinates, $c_{j,1}$ and $c_{k,1}$ can be found such that both $j$-th and $k$-th rows were modified, then the column embedding step does actually introduce three distortions.
We have implemented and executed a simulation of the embedding procedure described in this section which considers, among others, this particular case. For this reason the experimental results of the Kronecker product technique ("KP-technique (simulation)" in \figurename~\ref{fig:graphicKronecker}) have lower average distortion than the results obtained from the above $CI$-rate (plotted as "KP-technique" in \figurename~\ref{fig:graphicKronecker}).

\medskip

\section{Product of perfect $\zz$-linear codes}\label{sec:productz2z4}

The previous procedure deals with Hamming codes. Now, we will apply it to perfect $\zz$-linear codes. Let $\codi$ be a $\zz$-additive perfect code of type $(\alpha,\beta;\gamma,\delta)$ and binary length $n=2^m-1$, for $m\geq2$, and let $\HH_{\codi}$ be its parity check matrix. Take the code $\codi^{'}$ whose codewords are all the $n \times N$ matrices, where $N=2^{m-1}$, such that all rows are codewords in $\codi$, and so is the first column after applying the inverse of the extended Gray map $\Phi$.

Take blocks of $n \times N$ grayscale symbols in the source,

$$\begin{array}{ccc}
        x_{1,1},&\ldots,&x_{1,N}\\
	\vdots&\vdots&\vdots \\
	x_{n,1},&\ldots,&x_{n,N}
    \end{array}$$

where $N=2^{m-1}$, and translate them into $n$ vectors of $\alpha$ binary and $\beta$ quaternary coordinates, as stated in Section~\ref{sec:z2z4stegano} and explained in depth in~\cite{ririro}. At the same time, the first coordinate of those $n$ vectors is making up a binary vector of length $n$ which can also be seen as a vector of $\alpha$ binary and $\beta$ quaternary coordinates by means of the inverse of the extended Gray map $\Phi$ (see Section~\ref{sec:z2z4stegano}). Note that considering the $n$ rows and the first column, we end up having $n+1$ different vectors of binary length $n$.

The embedding procedure we will apply here is very similar to the KP-technique described in Section~\ref{sec:ppcGeneral}.
Once the $n+1$ vectors have been translated into $n+1$ vectors in $\Z_2^{\alpha} \times \Z_4^{\beta}$, we will proceed by steps: first, we will apply the embedding scheme from~\cite{ririro} to every row, and then we will apply it to the first column. Each distortion in the rows will involve adding or subtracting one unit to/from a grayscale symbol, and this requires considering the possibility of having extreme grayscale values problems. Recall that, unlike vectors in rows, the vector in the first column is only made up of the least significant bit of $n$ grayscale symbols and not of their two least significant bits. This means that any distortion over a coordinate in this vector will involve a flip in the least significant bit of a grayscale symbol $x_{i,1}$, for $i \in \{1,\ldots,n\}$, which leads us to conclude that, unlike the rows embedding step, no extreme grayscale values problem will ever crop up during the column embedding step.

Furthermore, as in the KP-technique from Section~\ref{sec:ppcGeneral}, during the column embedding step we have to consider different situations. The embedding method may require modifying a certain coordinate in the first column, and this coordinate may correspond to a row which was (or was not) modified during the row embedding step, or to a row that contains two distorted grayscale symbols, probably to deal with an extreme grayscale value problem. In any of these cases the action to be taken may vary, but still the aim is performing the appropriate changes in the affected row so that the distortion being introduced now by the column embedding step does not affect the embedding in the row. Take any two column vectors $\bh_j,\bh_k$ of order four in matrix $\HH_{\codi}$, such that one is the complementary of the other, that is $\bh_j=\bh_k+\dos$, where $\dos$ is the all-twos vector. The changes above mentioned will consist of considering that any distortion in coordinate $x_{i,1}$, for any $i \in \{1,\ldots,n\}$, can be compensated either by doing $x_{i,j-(\alpha+1)/2}+1$ and $x_{i,j-(\alpha+1)/2}+1$ or by doing $x_{i,j-(\alpha+1)/2}-1$ and $x_{i,j-(\alpha+1)/2}-1$. Note that, whenever possible, we will avoid modifying those grayscale symbols associated with column vectors in $\HH_{\codi}$ that are complementary of themselves, because in this case we would have to distort the associated symbol in two units instead of one, which would not conform to $\pm 1$-steganography.

\bigskip

By means of this method we can embed $m$ bits into the first column $x_{1,1},\ldots,x_{n,1}$ and also in every row of the block. Since the first column is made up of $n$ grayscale symbols and each row is made up of $N$ symbols, we are actually embedding $(n+1)m$ bits in $nN=n(n+1)/2$ symbols. It is easy to see that an upper bound of the average distortion for the symbols in the first column is $\frac{(n+3)/(n+1)}{(n+1)}$. As for the symbols in each row, the average distortion is given by $\frac{2N-1 + \frac{N-1}{2^{B-2}}}{2N^2} = \frac{2n+ \frac{n-1}{2^{B-2}} }{(n+1)^2}$ (see Section~\ref{sec:z2z4stegano}). Summing this up, an upper bound for the average distortion is
$$\frac{n \frac{(n+3)/(n+1)}{(n+1)} + \frac{n(n+1)}{2} \frac{2n+ \frac{n-1}{2^{B-2}} }{(n+1)^2} }{n(n+1)/2} =$$
$$\frac{ 2n+\frac{n-1}{2^{B-2}} }{(n+1)^2} \left ( \frac{n+3}{(n+ \frac{n-1}{2^{B-1}})(n+1)} + 1 \right ).$$

In a similar way as we did in Section~\ref{sec:ppcGeneral}, we can repeat this method over and over and generalize the computations of the average distortion and the embedding rate by taking the code whose codewords are all the $l$-dimensional matrices, where $l=n\times(n\times \cdots \times n \times N)$, such that their rows and the first component of every dimension are codewords in the $\zz$-additive perfect code $\codi$.

Let $D_l$ be the average distortion at the $l$-th step. For the first steps we have $D_1=\frac{2n+ \frac{n-1}{2^{B-2}}}{(n+1)^2}$ and $D_2=\frac{ 2n+\frac{n-1}{2^{B-2}} }{(n+1)^2}  \left ( \frac{n+3}{(n+ \frac{n-1}{2^{B-1}})(n+1)} + 1  \right )$. In the general case we have:
$$
D_l= \frac{2n+ \frac{n-1}{2^{B-2}} }{(n+1)^2} + \xi D_{l-1},
$$
where $\xi = \frac{n+3}{(n+ \frac{n-1}{2^{B-1}})(n+1)}$.

Now, the overall average distortion can be computed as $ \frac{2n+ \frac{n-1}{2^{B-2}} }{(n+1)^2}\left(1+\xi+\cdots +\xi^{l-1}\right)$, which converges asymptotically very fast to

$$\frac{2n+ \frac{n-1}{2^{B-2}} }{(n+1)^2}\left( \frac{1}{1-\xi} \right).$$

As for the embedding rate, it can be computed as $\frac{(1+n+n^2+\cdots +n^{l-1})m}{Nn^{l-1}}$, which converges to $\frac{mn}{N(n-1)}$.

Finally, we obtain a $CI$-rate of $\left ( \frac{2n+ \frac{n-1}{2^{B-2}} }{(n+1)^2} ( \frac{1}{1-\xi}) , \frac{mn}{N(n-1)} \right )$.

\medskip

\begin{figure}[htb!]
\centering 
\includegraphics[scale=0.5]{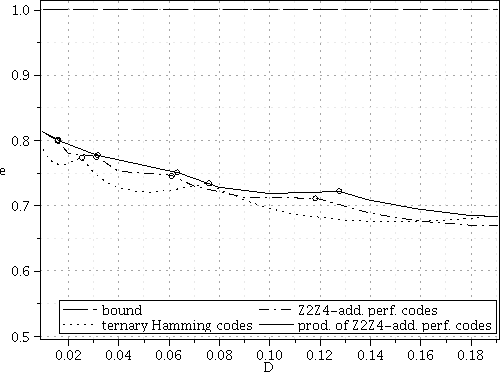}
\caption{Normalized embedding rate $e$ as a function of the average distortion $D$, of steganographic methods based on $\zz$-additive perfect codes~\cite{ririro} ("Z2Z4-add. perf. codes"), on ternary Hamming codes~\cite{fri},\cite{WiMa} and on the product of $\zz$-additive perfect codes ("prod. of Z2Z4-add. perf. codes").}\label{fig:graphicKroneckerZ2Z4}
\end{figure}

\figurename~\ref{fig:graphicKroneckerZ2Z4} shows a comparison of the normalized embedding rate $e=\frac{{H_q}^{-1}(E)}{D}$, for $q=3$, as a function of the average distortion $D$ for the steganographic method based on perfect $\zz$-linear codes~\cite{ririro}, the one based on ternary Hamming codes~\cite{fri},\cite{WiMa} and the new method based on the product of perfect $\zz$-linear codes. Recall that the distortion we have computed in the above $CI$-rate is an upper bound on the average distortion, meaning that lower distortion can be achieved in some particular cases, as it happened in the simulation results from Section~\ref{sec:ppcGeneral}.

\section{Conclusions}\label{sec:conclusions}

The use of perfect $\zz$-linear codes in $\pm 1$-steganography was first proposed in~\cite{ririro}. This method has a better performance compared to those based on the direct sum of ternary Hamming codes from~\cite{fri} and~\cite{WiMa}, and also deals with the extreme grayscale values more efficiently.

In this paper we have presented a technique based on products of these perfect $\zz$-linear codes. Therefore, the proposed method has all the advantages related to the performance and the processing of extreme grayscale values compared to the techniques based on the direct sum of ternary Hamming codes. Furthermore, we have shown that it performs better than the method in~\cite{ririro}.

\end{document}